\documentclass[]{article}
\usepackage{txfonts}
\usepackage{graphicx}
\usepackage{natbib}
\bibpunct{(}{)}{;}{a}{}{,} 

\newcommand{\hs}{\hspace{0.08cm}}

\begin{document}

\title{\bf Behavior of the reflection function of a plane--parallel 
           medium for directions of incidence and reflection 
           tending to horizontal directions}

\author{J. W. Hovenier \\ (J.W.Hovenier@uva.nl) \\
        Astronomical Institute 'Anton Pannekoek', \\
        University of Amsterdam, \\
        Kruislaan 403, 1098 SJ Amsterdam, The Netherlands \\
        \and
	D. M. Stam \\
        DEOS, Department of Aerospace Engineering, \\
        TU Delft, Kluyverweg 1, 2629 HS, Delft, The Netherlands \\
        SRON Netherlands Institute for Space Research, \\ 
        Sorbonnelaan 2, 3584 CA Utrecht, The Netherlands}

\maketitle

                                                                                
                                                                                
\abstract{The atmospheres of (exo) planets and moons, as well as reflection 
nebulae, contain in general independently scattering particles in random 
orientation and are often supposed to be plane-parallel. 
Relations are presented for the (bidirectional) reflection function and
several related functions of such a medium 
in case the directions of incidence and reflection both 
tend to horizontal directions. 
The results are quite general. The medium may be semi-infinite or finite, 
with or without a 
reflecting surface underneath, and vertically homogeneous or 
inhomogeneous. Some approximative formulae for the reflection function 
of a plane-parallel medium with independently scattering particles in 
random orientation, including Lambert's law, may be very 
inaccurate if the directions of incidence and reflection are both nearly 
horizontal. \\
          
\noindent keywords: planets and satellites -
          scattering -
          radiative transfer -
          reflection nebulae}
                                                                                
                                                                                
\section{Introduction}

A well--known subject in astrophysics concerns multiple scattering of 
(electromagnetic) radiation in an extended medium containing small, 
independently scattering particles 
\citep[see e.g.][]{Ambarzumian1943,1950ratr.book.....C,
1975QB603.A85S6213,1980vandeHulst,1974SSRv...16..527H,
2004Hovenier,2006Mishchenko}.
Examples of such 
media are provided by the atmospheres of (exo)planets and satellites, 
as well as reflection nebulae and protoplanetary disks. The medium is 
often supposed to be locally plane--parallel, so that one can use the 
theory developed for a plane--parallel atmosphere, i.e.\ a horizontally 
homogeneous atmosphere of infinite horizontal extent. The radiation 
(which we will also call light) coming from a distant source, like the 
Sun or a star, may illuminate the top of the atmosphere and then be 
scattered by the particles inside before leaving the atmosphere at 
the top in all upward directions. This is called reflected radiation 
and, neglecting polarization, the angular distribution of its (specific) 
intensity can be expressed by means of the so--called reflection function. 

The reflection function is an important fundamental property of an
atmosphere, normalized so that it is identically equal to one for a 
perfectly white surface following Lambert's law. Once the reflection 
function of an atmosphere has been obtained, one readily finds the 
angular distribution of the intensity of the reflected radiation for 
any angular distribution of incident radiation at the top. In the 
literature this function has a variety of names, such as reflection 
coefficient 
\citep[][]{1975QB603.A85S6213,1997Yan},
bidirectional reflection function 
\citep[][]{1999JQSRT..63..409M} and reflectance factor
\citep[][]{1993tres.book.....H}.

Numerous theoretical and numerical studies of reflection functions have 
been reported in books and papers. Yet, very little attention was given 
to the limiting case when the directions of incident and reflected 
radiation both tend to horizontal directions. This case not only provides 
more insight into the angular distribution of the reflected radiation, but 
it is also important for exact and approximate computations, in particular 
when discontinuities are involved. This was shown for the intensity of 
the reflected radiation, first when polarization is neglected 
\citep[][]{2006JQSRT.101....1H} and later when it is taken into account 
\citep[][]{2007JQSRT.107...83H}. As far as the reflection function is concerned, 
an interesting statement was made by 
\citet{1980vandeHulst}, namely that the 
reflection function becomes infinitely large for horizontal directions 
of both incidence and reflection. But he restricted himself to the 
azimuth--independent terms in a Fourier series expansion and he did not 
give any evidence or clarification regarding this statement.

The principal aim of this paper is to present a comprehensive treatment of 
the behavior of the reflection function and related functions when the 
directions of incidence and reflection both tend to horizontal directions. 
The organization of this paper is as follows. Some basic concepts and 
definitions are discussed in Sect.~\ref{sect2}. In Sect.~\ref{sect3}, 
the limiting process of directions of incidence and reflection tending to 
horizontal directions, while keeping the azimuth difference of the 
directions fixed, is considered for, respectively, the reflected intensity,
the reflection function and orders of scattering of both. Simple examples 
are given in Sect.~\ref{sect4} for vertically homogeneous as well as 
inhomogeneous media. Section~\ref{sect5} is devoted to functions that 
are related to the reflection function. The azimuth dependence of the 
incident and reflected light is treated in Sect.~\ref{sect6}. 
Approximations are discussed in Sect.~\ref{sect7} and some concluding 
remarks are presented in Sect.~\ref{sect8}.
                         
\section{Concepts and defintions}
\label{sect2}

We consider a plane-parallel atmosphere composed of randomly oriented 
particles, which may include gas molecules. The particles scatter 
radiation independently and without change of wavelength in all 
directions with a scattering angle distribution called the phase function. 
Since we are mainly interested in natural particles and physically realistic 
model particles we will assume that the albedo of single scattering is 
positive but not larger than one and the phase function is a positive, 
bounded and continuous function of directions. It is normalized so that 
its average over all directions equals unity.

There are no internal sources in the atmosphere. A parallel beam of radiation, 
coming from a distant source, is incident on each point of the top of the 
atmosphere. The net flux per unit area normal to this beam is $\pi F_0$. 
The direction of the incident beam is given by $\mu_0$, which is the cosine 
of the angle this direction makes with the downward normal, and an azimuthal 
angle $\phi_0$. 
Polarization is ignored and we focus on the reflected radiation, i.e.\ 
the radiation that emerges at the top of the atmosphere. Its direction 
is described by $\mu$, the cosine of the angle this direction makes with 
the upward normal, and an azimuthal angle $\phi$. The azimuthal angles 
are measured from an arbitrary zero direction in an arbitrary sense  
and only the difference, $\phi-\phi_0$, is relevant, since the medium is 
horizontally homogeneous. It should be noted that with our definitions $\mu$ 
and $\mu_0$ are non--negative and we have $0 \leq \phi - \phi_0 \leq 2 \pi$. 
In this paper we are mainly interested in the question what happens with the 
reflection properties of the atmosphere if $\mu$ and $\mu_0$ both approach 
zero, starting from values larger than zero.

The intensity of the reflected radiation can be written in the form
\begin{equation}
   I^{\rm t}(\mu,\mu_0,\phi-\phi_0) = \mu_0 R(\mu,\mu_0,\phi-\phi_0) F_0.
\label{eq1}
\end{equation}
Here and hereafter the superscript ${\rm t}$ is used to indicate the top of the 
atmosphere and $R(\mu,\mu_0,\phi-\phi_0)$ is the reflection function. The 
intensity at the top and the reflection function are both nonnegative. 
Once the reflection function has been obtained one readily finds the 
intensity of the reflected radiation for a parallel beam of incident 
radiation by using Eq.~\ref{eq1}, while for multi--directional incident 
light an integration over all incident directions must be performed.

\section{Approaching the origin in the ($\mu_0,\mu$)-plane}
\label{sect3}

A function of one real variable may have two different limits at a point 
of the real number axis, namely a right--hand limit and a left--hand limit. 
For a function of two or more real variables there are many more 
possibilities to approach a point in the relevant multi--dimensional 
space and this may or may not correspond to a number of different values 
to which the function approaches 
\citep[][]{Courant1962}.
This will be considered 
in the following sections for, respectively, the intensity of the reflected 
radiation, the reflection function and orders of scattering of both.

\subsection{Intensity of the reflected  radiation}
\label{sect3.1}

\begin{figure}
\centering
\resizebox{7cm}{!}{\includegraphics{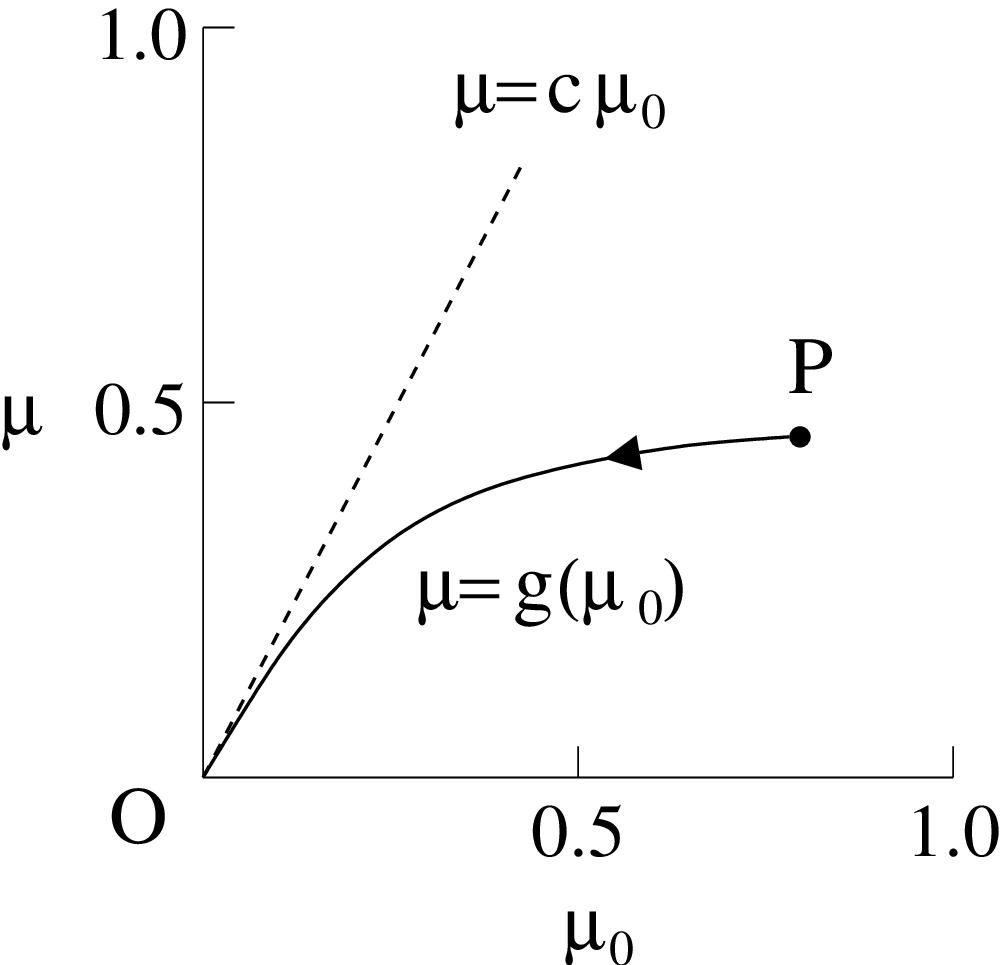}}
\caption{A point $P$ approaches the origin $O$ of a Cartesian coordinate 
         system along a curve (solid line) represented by $\mu= g(\mu_0)$, 
         which has a slope $c$ at $O$ with respect to the positive 
         $\mu_0$-axis. The tangent of the curve at $O$ has also been drawn 
         (dashed line).}
\label{fig1}
\end{figure}

Let us keep the azimuth difference fixed so that the functions 
$I^{\rm t}(\mu,\mu_0,\phi-\phi_0)$ and $R(\mu,\mu_0,\phi-\phi_0)$ are 
functions of the two variables, $\mu$ and $\mu_0$. We can approach 
the point $\mu = \mu_0 = 0$ in various ways. This can be visualized 
by saying that we can approach the origin, $O$, in the first 
quadrant of a Cartesian ($\mu_0,\mu$)-coordinate system by following 
different paths (curves, including straight lines). Suppose we represent 
such a curve by means of a continuous function $\mu = g(\mu_0)$ through 
$O$ with a definite tangent at $O$ 
(see Fig.~\ref{fig1}). If we now approach the origin along this 
curve, the ratio $g(\mu_0)/\mu_0$ will tend to the slope of the tangent 
at $O$, which we denote as $c$. For a non--perpendicular tangent at 
$O$ the value of $c$ is finite and equals the right--hand derivative 
of $g(\mu_0)$ at $O$. Evidently, $\mu_0$ and $g(\mu_0)$ are both 
non-negative, so that $c$ is non-negative. If the curve is a straight line 
we simply have $g(\mu_0) = c \mu_0$. If we first let $\mu$ and then 
$\mu_0$ tend to zero we have $c= 0$. If we first let $\mu_0$ and then 
$\mu$ approach zero we can treat this case separately or by letting 
$c$ tend to infinity.

As an illustration we may consider the special case of isotropic 
scattering in a semi--infinite homogeneous atmosphere with an albedo 
of single scattering $a$. The phase function is then identically equal 
to one and we have \citep[][]{1950ratr.book.....C}
\begin{equation}
   I^{\rm t}_1(\mu,\mu_0,\phi-\phi_0) = \frac{a F_0}{4} \frac{\mu_0}{\mu+\mu_0},
\label{eq2}
\end{equation}
where the subscript 1 refers to the first order of scattering. Writing 
$\lim_{\mu,\mu_0 \rightarrow 0}$ for the limit if we approach $O$ along 
a curve represented by $\mu = g(\mu_0)$ we readily find
\begin{equation}
   \lim_{\mu,\mu_0 \rightarrow 0} I^{\rm t}_1 (\mu,\mu_0,\phi-\phi_0) = \frac{a F_0}{4(c+1)},
\label{eq3}
\end{equation}
which shows that the result depends on $c$, i.e.\ on the path that is 
taken to approach the origin. Consequently, there is a discontinuity for 
the intensity of the reflected radiation when the directions of incidence 
and reflection both become horizontal. This was called a peculiar discontinuity 
by \citet{2006JQSRT.101....1H}, since it looks at first glance rather surprising. 

In general, the scattering may be anisotropic, the atmosphere may be 
vertically inhomogeneous and its optical thickness may be finite with a 
reflecting or totally absorbing surface underneath the atmosphere. The 
intensity of the reflected radiation in this general case for near--horizontal 
directions was also considered by \citet{2006JQSRT.101....1H}. They found 
a peculiar discontinuity in this intensity when $\mu$ and $\mu_0$ both 
approach the origin in the ($\mu_0,\mu$)-plane, which can be written as
\begin{eqnarray}
   \lim_{\mu,\mu_0 \rightarrow 0} I^{\rm t}(\mu,\mu_0,\phi-\phi_0) & = &
   \lim_{\mu,\mu_0 \rightarrow 0} I^{\rm t}_1(\mu,\mu_0,\phi-\phi_0) \nonumber \\ 
   & = & \frac{a^{\rm t}}{4(c+1)}  Z^{\rm t}(\cos (\phi-\phi_0)) F_0, 
\label{eq4}
\end{eqnarray}
where $a^{\rm t}$ is the albedo of single scattering at the top of the atmosphere 
and $Z^{\rm t}(\cos \Theta)$ is the phase function at the top of the atmosphere 
with scattering angle $\Theta$. Naturally, in Eq.~\ref{eq4} the same path must be 
followed for the two limits (see Fig.~\ref{fig1}). Consequently, the 
following conclusions can be drawn for the intensity of the reflected 
radiation in the limit of $\mu$ and $\mu_0$ both being zero:
a) the optical thickness of the atmosphere and the reflection properties 
of the underlying surface are irrelevant,
b) the values of the albedo of single scattering and the phase function 
need only to be known at the top of the atmosphere,
c) orders of scattering higher than the first do not contribute,
d) for any path with $c$ unequal to infinity the azimuth dependence is 
proportional to the scattering angle dependence of the phase function at 
the top. The constant of proportionality becomes zero if $c$ equals infinity. 
A similar statement was made by \citet{1935Phy.....2..363M} for a 
semi-infinite homogeneous 
atmosphere, but he did not provide a correct proof and did not mention that 
the constant of proportionality depends on the way $\mu$ and $\mu_0$ tend 
to zero. For more details about the behavior of the  reflected intensity 
for directions of incidence and reflection that both tend to horizontal 
directions we refer to \citet{2006JQSRT.101....1H}.

\subsection{The reflection function}
\label{sect3.2}

Let us now consider what happens with the reflection function on 
approaching the origin in the ($\mu_0,\mu$)-plane. Combining 
Eqs.~\ref{eq1} and~\ref{eq4} gives
\begin{eqnarray}
   \lim_{\mu, \mu_0 \rightarrow 0} \hs \mu_0 R(\mu,\mu_0,\phi-\phi_0) & = & \nonumber \\ 
   \lim_{\mu, \mu_0 \rightarrow 0} \hs \mu_0 R_1(\mu,\mu_0,\phi-\phi_0) & = &
   \frac{a^{\rm t}}{4(c+1)} Z^{\rm t}(\cos (\phi-\phi_0)).
\label{eq5}
\end{eqnarray}

Using the principle of reciprocity \citep[][]{1980vandeHulst} we have for all 
orders of scattering and their sum
\begin{equation}
  R(\mu,\mu_0,\phi-\phi_0)= R(\mu_0,\mu,\phi-\phi_0).
\label{eq6}
\end{equation}
So, interchanging $\mu$ and $\mu_0$ in Eq.~\ref{eq5} and taking limits 
following the same paths as before we find
\begin{eqnarray}
   \lim_{\mu,\mu_0 \rightarrow 0} \hs \mu R(\mu,\mu_0,\phi-\phi_0) & = & \nonumber \\
   \lim_{\mu,\mu_0 \rightarrow 0} \hs \mu R_1(\mu,\mu_0,\phi-\phi_0) 
   & = & \frac{a^{\rm t}c}{4(c+1)} Z^{\rm t}(\cos (\phi-\phi_0)), 
\label{eq7}
\end{eqnarray}
since $c$ in Eq.~\ref{eq5} had to be replaced by $1/c$, i.e.\ the 
slope of $g(\mu_0)$ at $O$ with the positive $\mu$-axis. 
Since $c$ is nonnegative, the limits in Eqs.~\ref{eq5} and~\ref{eq7} are bounded.
So it follows from either one that
\begin{equation}
   \lim_{\mu,\mu_0 \rightarrow 0} \mu \mu_0 R(\mu,\mu_0,\phi-\phi_0)= 0.
\label{eq8}
\end{equation}
Since the right-hand side of this equation does not depend on $c$, this 
limit is the same for all curves represented by $\mu = g(\mu_0)$. More 
limits of this type, i.e.\ not depending on $c$, will be encountered 
further down in this paper. 

By adding Eqs.~\ref{eq5} and~\ref{eq7}, we obtain
\begin{eqnarray}
   \lim_{\mu,\mu_0 \rightarrow 0} (\mu + \mu_0) R(\mu,\mu_0,\phi-\phi_0) & = & \nonumber \\
   \lim_{\mu,\mu_0 \rightarrow 0} (\mu + \mu_0) R_1(\mu,\mu_0,\phi-\phi_0) 
   & = & \frac{a^{\rm t}}{4} Z^{\rm t}(\cos (\phi-\phi_0)). 
\label{eq9}
\end{eqnarray}
This result has been reported for the special case of a semi-infinite, 
homogeneous atmosphere by several authors
\citep[see e.g.][]{1975QB603.A85S6213,2006Mishchenko,2001Kok}, 
but without a rigorous proof.

Since the far right-hand side of Eq.~\ref{eq9} is positive we must have
\begin{equation}
   \lim_{\mu,\mu_0 \rightarrow 0} R_1(\mu,\mu_0,\phi-\phi_0)= \infty
\label{eq10}
\end{equation}
and 
\begin{equation}
   \lim_{\mu,\mu_0 \rightarrow 0} R(\mu,\mu_0,\phi-\phi_0)= \infty, 
\label{eq11}
\end{equation}
because zero or any finite number for these limits would be in conflict with 
Eq.~\ref{eq9}. Consequently, we have proved that $R_1(\mu,\mu_0,\phi-\phi_0)$ as 
well as $R(\mu,\mu_0,\phi-\phi_0)$ have a discontinuity if $\mu$ and 
$\mu_0$ are both zero. The nature of these discontinuities is, however, 
quite different from the peculiar discontinuities for the intensities 
discussed in Sect.~\ref{sect3.1}.

\subsection{Orders of scattering}
\label{sect3.3}

The reflected intensity and the reflection function can be written as 
a sum (series) of nonnegative terms representing orders of scattering 
\citep[][]{1980vandeHulst}.
In view of Eq.~\ref{eq4} we have for the $n$-th order of scattering
\begin{equation}
   \lim_{\mu,\mu_0 \rightarrow 0} I^{\rm t}_n(\mu,\mu_0,\phi-\phi_0) = 0
   \hspace{0.3cm} \mbox{for} \hspace{0.3cm} n > 1.
\label{eq12}
\end{equation}
However, we cannot infer from Eqs.~\ref{eq10}-\ref{eq11} what will happen 
with $R_n(\mu,\mu_0,\phi-\phi_0)$ for $n > 1$ if $\mu$ and $\mu_0$ 
both approach zero. The result might be zero, a finite positive number 
or infinity. However, some interesting properties for the higher orders 
of scattering of the reflection function can be obtained as follows. 
Equation~\ref{eq9} shows that
\begin{equation}
   \lim_{\mu,\mu_0 \rightarrow 0} \hs (\mu+\mu_0) R_n(\mu,\mu_0,\phi-\phi_0) = 0 
   \hspace{0.3cm} \mbox{for} \hspace{0.3cm} n > 1.
\label{eq13}
\end{equation}
Furthermore, writing
\begin{equation}
   \frac{R(\mu,\mu_0,\phi-\phi_0)}{R_1(\mu,\mu_0,\phi-\phi_0)} = 
   \frac{(\mu+\mu_0) R(\mu,\mu_0,\phi-\phi_0)}{(\mu+\mu_0) R_1(\mu,\mu_0,\phi-\phi_0)}
\label{eq14}
\end{equation}
and using Eq.~\ref{eq9} we find                                   
\begin{equation}
   \lim_{\mu,\mu_0 \rightarrow 0} 
   R(\mu,\mu_0,\phi-\phi_0)/R_1(\mu,\mu_0,\phi-\phi_0) = 1,
\label{eq15}
\end{equation}
since the right hand side of Eq.~\ref{eq9} is positive. Similarly, we obtain
\begin{equation}
   \lim_{\mu,\mu_0 \rightarrow 0} 
   R_n(\mu,\mu_0,\phi-\phi_0)/R_1(\mu,\mu_0,\phi-\phi_0) = 0,
   \hspace{0.3cm} \mbox{for} \hspace{0.3cm} n > 1 
\label{eq16}
\end{equation}
The ratio of the reflection function to its first order term is an important function, 
since it is the correction factor to be applied to the easily computed first 
order term to obtain the reflection function. Numerical studies of this correction 
factor for homogeneous atmospheres by \citet{1980vandeHulst} have shown that for 
isotropic scattering  the correction factor has a maximum  of 8.455, but 
that it becomes much smaller for nearly horizontal directions of incidence and 
reflection. He also found a similar behavior for anisotropic phase functions 
and mentioned Eq.~\ref{eq15}. 

\section{Examples}
\label{sect4}

To illustrate and check the results of the preceding section we will now 
discuss some simple examples. First for homogeneous and then for 
inhomogeneous atmospheres.

\subsection{Homogeneous atmospheres}
\label{sect4.1}

Let us consider again the special case of isotropic scattering in a 
semi-infinite, homogeneous atmosphere with albedo of single scattering $a$.
This implies that there is no azimuth dependence for the reflected 
radiation, so that we can omit $\phi-\phi_0$ in equations. The first 
two orders of scattering of the reflection function can readily be 
computed by analytic integration over optical depth 
\citep[][]{1971A&A....13....7H} or by iteration of an invariance relation 
\citep[][]{Ambarzumian1943,2006Mishchenko}. The results are as follows
\begin{equation}
   R_1(\mu,\mu_0) = \frac{a}{4(\mu+\mu_0)} 
\label{eq17}
\end{equation}
and
\begin{equation}
   R_2(\mu,\mu_0) = \frac{a^2}{8(\mu+\mu_0)} (k(\mu) + k(\mu_0)),
\label{eq18}
\end{equation}
where
\begin{equation}
   k(\mu)= \mu  \ln (1 + 1/\mu). 
\label{eq19}
\end{equation}

For the sum over all orders we have \citep[][]{1950ratr.book.....C} 
\begin{equation}
   R(\mu,\mu_0)= \frac{a}{4(\mu + \mu_0)} H(\mu) H(\mu_0),
\label{eq20}
\end{equation}
where $H(\mu)$ is a well-known function, depending on $a$, with 
$H(0)=1$ \citep[][]{1960busbridge.book}. Equations~\ref{eq17}-\ref{eq20} show 
that the reflection function and its first two orders of scattering 
tend to infinity if $\mu$ and $\mu_0$ both approach zero. 
Consequently, this also holds for the multiple scattering
component of the reflection function.
Clearly 
Eqs.~\ref{eq17}-\ref{eq20} are in agreement with Eqs.~\ref{eq5}-\ref{eq11}
and~\ref{eq13}-\ref{eq16},
since $k(0)= 0$. Figure~\ref{fig2} shows $R_1(\mu,\mu_0)$ 
and $R(\mu,\mu_0)$ as functions of $\mu$ if $a=1$ and $\mu= \mu_0$. 
Both functions are seen to be strongly increasing when the directions of 
incidence and reflection tend to horizontal directions.

It follows from Eqs.~\ref{eq17} and~\ref{eq20} that the correction factor
\begin{equation}
   R(\mu,\mu_0)/R_1(\mu,\mu_0) = H(\mu) H(\mu_0).
\label{eq21}
\end{equation}
This factor varies between 1 (if $\mu=\mu_0=0$ and $a$ is arbitrary) 
and 8.455 (if $\mu=\mu_0=1$ and $a=1$). 
Fig.~\ref{fig2} also shows this correction factor in case $\mu=\mu_0$.

A lesson to be learned from this simple case is that one should not 
assume that the discontinuity of the reflection function for horizontal 
directions of incidence and reflection will always disappear upon 
integration. Indeed we find from Eq.~\ref{eq17} 
\begin{equation}
   \int_{0}^{1} d \mu_0 \hs R_1(\mu,\mu_0) = \frac{a}{4} \ln(1+1/\mu),
\label{eq22}
\end{equation}
which tends to infinity if $\mu$ tends to 0. So this approach to 
infinity must also hold if the reflection function is integrated 
in the same way, since the sum of all orders of the reflection 
function cannot be smaller than the first order only. Another 
way to prove this can be obtained from the definition of the $H$-function 
written in the form
\begin{equation}
   \int_{0}^{1} d \mu_0 \hs R(\mu,\mu_0) = \frac{1}{2 \mu} (H(\mu) - 1) 
\label{eq23}
\end{equation}
and the behavior of the $H$-function when $\mu$ approaches zero 
\citep[][]{1980vandeHulst}.

Explicit expressions for the reflection function of a semi-infinite 
or finite homogeneous atmosphere on top of a black surface in terms of 
functions $H(\mu)$, $X(\mu)$ and $Y(\mu)$ have been published 
\citep[see e.g.][]{1950ratr.book.....C}
for phase functions that can be written as a sum of 
a few Legendre polynomials, including Rayleigh scattering. Since
$H(0)= 1$, $X(0)= 1$ and $Y(0)= 0$ \citep[][]{1960busbridge.book} it can readily be 
verified that in all these cases Eqs.~\ref{eq5}--\ref{eq11} and 
Eq.~\ref{eq15} are valid. 

\subsection{Inhomogeneous atmospheres}
\label{sect4.2}

In numerical calculations a vertically inhomogeneous atmosphere is often 
modeled as a stack of homogeneous layers. Then only the albedo of single 
scattering and the phase function of the top layer are relevant for the 
reflected intensity when $\mu$ and $\mu_0$ both approach zero. When these 
values are substituted in Eqs.~\ref{eq4},~\ref{eq5},~\ref{eq7}, and~\ref{eq9},
Eqs.~\ref{eq4}-\ref{eq16} are also valid for a stack of homogeneous layers.

Another example of scattering in an inhomogeneous atmosphere is provided 
by isotropic scattering in a semi-infinite atmosphere with albedo of 
single scattering given by
\begin{equation}
   a(\tau)= \exp(-\tau),
\label{eq24}
\end{equation}
where $\tau$ is the optical depth measured from the top of the atmosphere 
downwards. In this case we have for $\mu_0 > 0$ \citep[][]{2006JQSRT.101....1H}
\begin{equation}
   I^{\rm t}_1(\mu,\mu_0) = \frac{F_0}{4} \frac{\mu_0}{(\mu + \mu_0 + \mu \mu_0)}
\label{eq25}
\end{equation}
so that 
\begin{equation}
   R_1(\mu,\mu_0) = \frac{1}{4} \frac{1}{(\mu + \mu_0 + \mu \mu_0)}.
\label{eq26}
\end{equation}
It should be noted that the denominators in the last two equations are not 
simply the sum of $\mu$ and $\mu_0$. Eqs.~\ref{eq25}--\ref{eq26} and the 
expressions reported by \citet{1997Yan} for the total intensity of the 
reflected radiation and the reflection function are in agreement with 
Eqs.~\ref{eq4}-\ref{eq16}. So the latter equations provide a useful check.

\section{Related functions}
\label{sect5}

In addition to the reflection function several alternative functions 
for describing the reflective properties of a plane-parallel atmosphere
are found in the literature. If the difference with the reflection 
function is only a constant factor all preceding equations must be 
translated by simply taking this constant into account. But the 
difference may also involve a function of $\mu$ and/or $\mu_0$ and then 
the translation is less trivial. In such cases one can derive expressions 
for the alternative function by starting with an expression for the 
intensity of the reflected light in terms of the alternative function 
and then working along similar lines as we did for the reflection function 
or directly use relations for the reflection function to get corresponding 
results for the alternative function. We give some examples.  

\citet{1950ratr.book.....C} defined what he called the scattering 
function by writing
\begin{equation}
   I^{\rm t}(\mu,\mu_0,\phi-\phi_0) = \frac{F_0}{4 \mu} S(\mu,\phi;\mu_0,\phi_0),
\label{eq27}
\end{equation}
so that in view of Eq.~\ref{eq1} the scattering function is related to 
the reflection function as follows
\begin{equation}
   S(\mu,\phi;\mu_0,\phi_0)= 4 \hs \mu \mu_0 \hs R(\mu,\mu_0,\phi-\phi_0).
\label{eq28}
\end{equation}
It follows immediately from this equation and Eq.~\ref{eq8} that
\begin{equation}
   \lim_{\mu,\mu_0 \rightarrow 0} S(\mu,\phi;\mu_0,\phi_0)= 0. 
\label{eq29}
\end{equation}
So the same must be true for all orders of scattering, since we are 
dealing here with a sum of nonnegative numbers. 
Consequently, the orders of scattering of the scattering function, 
nor the function itself, tend to infinity if we approach the origin 
of the ($\mu_0,\mu$)-plane along a path given by $\mu= g(\mu_0)$. 
Using Eq.~\ref{eq9} we find 
\begin{eqnarray}
   \lim_{\mu,\mu_0 \rightarrow 0} (1/\mu + 1/\mu_0) S(\mu,\phi;\mu_0,\phi_0) & = & \nonumber \\
   \lim_{\mu,\mu_0 \rightarrow 0} (1/\mu + 1/\mu_0) S_1(\mu,\phi;\mu_0,\phi_0) & = &  
   a^{\rm t} Z^{\rm t}(\cos (\phi-\phi_0)), 
\label{eq30}
\end{eqnarray}
which again shows the special role played by single scattering. 

Some authors still use the scattering function, but the reflection 
function is more widely used today. Each one has certain advantages 
and disadvantages. \citet{1993tres.book.....H} defined several alternatives for 
the reflection function, most of which differ only by a constant 
factor from the reflection function. But his so-called bidirectional 
reflection function $r(\mu,\mu_0,\phi-\phi_0)$ has the same 
peculiar discontinuity as our $I^{\rm t}(\mu,\mu_0,\phi-\phi_0)/(\pi F_0)$.

\section{Azimuth dependence}
\label{sect6}

So far we have kept the azimuth difference $\phi-\phi_0$ fixed and 
considered what happens when not only the incident radiation tends 
to the horizontal direction in the $\phi_0$-plane but also the reflected 
radiation tends to the horizontal direction in the $\phi$-plane. 
More generally, the azimuthal angles may change when $\mu$ and $\mu_0$ 
tend to zero, i.e.\ ($\mu,\phi$) may tend to $(0,\bar{\phi})$ and 
($\mu_0,\phi_0$) to ($0,\bar{\phi_0}$) in some way. Clearly, the 
discontinuities for a plane-parallel atmosphere will then remain and 
we must only replace $\cos(\phi-\phi_0)$ by $\cos(\bar{\phi} - \bar{\phi_0})$ 
in Eqs.~\ref{eq4},~\ref{eq5},~\ref{eq7},and~\ref{eq9}. 
Such situations generally occur in 
spectrophotometry of planets and satellites when one moves the line of 
sight to the intensity poles ($\mu = \mu_0 = 0$), along the limb $(\mu=0)$, 
the terminator $(\mu_0=0)$ or an intermediate intensity meridian 
$(\mu= c \mu_0)$ with finite $c > 0$. This must be realized in 
computations when the outer layers are modeled as a plane-parallel 
atmosphere \citep[][]{2006JQSRT.101....1H}.

The azimuth dependence of intensities and reflection functions are 
often handled by making Fourier series expansions, e.g.\ by writing 
for an arbitrary function
\begin{equation}
  f(\mu,\mu_0, \phi-\phi_0) =
  f^0(\mu,\mu_0) + 2 \Sigma_{m=1}^{\infty} f^m(\mu,\mu_0) \cos(m(\phi-\phi_0)),
\label{eq31}
\end{equation}
where the upper index $m$ denotes the Fourier index. In this way we 
readily find expressions for each Fourier component separately from the 
equations given in preceding sections. 
For instance, writing $Z^{\rm t}(0,0,\phi-\phi_0)$ for
the phase function at the top of the atmosphere in case $\mu= \mu_0= 0$, we
obtain for $m \geq 0$ using Eqs.~\ref{eq5} and~\ref{eq7}
\begin{equation}
   \lim_{\mu,\mu_0 \rightarrow 0} \mu_0 \hs R^m(\mu,\mu_0) = 
   \frac{a^{\rm t}}{4(c+1)} Z^{{\rm t,}m}(0,0)  
\label{eq32}
\end{equation}
and  
\begin{equation}
   \lim_{\mu,\mu_0 \rightarrow 0} \mu \hs R^m(\mu,\mu_0) = 
   \frac{a^{\rm t} c}{4(c+1)} Z^{{\rm t,}m}(0,0).
\label{eq33}
\end{equation}
Adding the last two equations gives
\begin{equation}
   \lim_{\mu,\mu_0 \rightarrow 0} (\mu + \mu_0) R^m(\mu,\mu_0) = 
   \frac{a^{\rm t}}{4} Z^{{\rm t,}m}(0,0).
\label{eq34}
\end{equation}

The azimuthal average of the phase function is positive, since we have 
assumed that the phase function itself is positive. Hence, 
the $m=0$ component of the phase function is positive and it follows 
from Eq.~\ref{eq34} that 
\begin{equation}
   \lim_{\mu,\mu_0 \rightarrow 0} R^0(\mu,\mu_0) = \infty.
\label{eq35}
\end{equation}
However, $Z^{{\rm t,}m}(0,0)$ can be positive, negative or zero if $m > 0$. 
Consequently, if $\mu$ as well as $\mu_0$ tend to zero the limit of 
$R^m(\mu,\mu_0)$ may be plus infinity, minus infinity or it may be 
impossible to determine this limit from Eq.~\ref{eq34}. 
This may be illustrated by considering the following simple cases.

\noindent (i) The phase function
\begin{equation}
   Z^{\rm t}(\cos \Theta) = 1 - \frac{1}{2} \cos \Theta
\label{eq36}
\end{equation}
has 
\begin{equation}
   Z^{{\rm t,}0}(0,0)= 1
\label{eq37}
\end{equation}
and 
\begin{equation}
   Z^{{\rm t,}1}(0,0)= -1/4.
\label{eq38}
\end{equation}

\noindent (ii) Rayleigh's phase function, i.e.\
\begin{equation}
   Z^{\rm t}(\cos \Theta)= \frac{3}{4} (1+\cos^2 \Theta)
\label{eq39}
\end{equation}
has
\begin{eqnarray}
   Z^{{\rm t,}0}(0,0) & = &  9/8 \\
   Z^{{\rm t,}1}(0,0) & = & 0 \\
   Z^{{\rm t,}2}(0,0) & = & 3/16. 
\label{eq40}
\end{eqnarray}
Using the explicit expressions \citep[][]{1950ratr.book.....C} for the reflected 
intensity of a homogeneous semi-infinite atmosphere with $a=1$ for 
these two phase functions one can easily find that Eqs.~\ref{eq32}--\ref{eq34}
are satisfied.         

\section{Approximations}
\label{sect7}

Accurate computations of the reflection function of a plane--parallel 
atmosphere are generally not easy and also laborious, especially when 
results are needed for many model parameters. Therefore, it is not 
surprising that a variety of approximative formulae have been proposed 
\citep[][]{1975QB603.A85S6213,1980vandeHulst,2001Kok}.
Here we wish to point 
out that at least some of these are not realistic for directions of 
incidence and reflection that are both nearly horizontal. 

For instance, according to the popular Lambert's reflection law the reflection 
function is a constant not larger than one for all directions of incidence and 
reflection. So this reflection function does not obey Eq.~\ref{eq9} and can never 
tend to infinity, which is in conflict with Eq.~\ref{eq11}. 

Another example is the so--called "rapid-guess formula" of \citet{1980vandeHulst} 
for a non-absorbing homogeneous semi-infinite atmosphere
\begin{equation}
   R^0(\mu,\mu_0) = 1 + p(1-3 \mu/2) (1-3 \mu_0/2),
\label{eq41}
\end{equation}
where the upper index, $0$, refers to the azimuth independent term and $p$ 
is a finite constant. This formula clearly violates Eq.~\ref{eq35} and, therefore, 
may give unacceptable errors for nearly horizontal directions of incidence and 
reflection. To show this we consider isotropic scattering for which the 
value $p=0.4$ was recommended by van de Hulst (1980). Using this value in 
the rapid--guess formula with $\mu= \mu_0= 0.1$ gives for the reflection 
function 1.289, which is 33.72~$\%$ too low, since accurate multiple 
scattering calculations give 1.94485. For smaller values of $\mu$ and 
$\mu_0$ the errors in the reflection function are still larger. Integration 
of the reflection function as in Eq.~\ref{eq23} gives for $\mu= 0.1$ the accurate 
value of 1.23675, but Eq.~\ref{eq41} gives 1.0850 which is still 12.27~$\%$ too low.

Finally, we mention a formula that is often used for a homogeneous atmosphere 
above a black surface when the optical thickness $b$ is a small positive 
number, namely
\begin{equation}
   R(\mu, \mu_0, \phi-\phi_0)= \frac{a b}{4 \mu \mu_0} Z(\cos \Theta),
\label{eq42}
\end{equation}
where
\begin{equation}
   \cos \Theta= -\mu \mu_0 + \sqrt{(1-\mu^2)(1-\mu_0^2)} \cos(\phi-\phi_0).
\label{eq43}
\end{equation}
Although Eq.~\ref{eq42} does not violate Eq.~\ref{eq11} the approach of this 
approximate 
reflection function to infinity if $\mu$ and $\mu_0$ both tend to zero
is apparently not correct, since Eqs.~\ref{eq5}  
and~\ref{eq7}-\ref{eq9} are not satisfied.

Consequently, if $\mu$ and $\mu_0$ are very small one should be careful with 
using approximations for the reflection function. Fortunately, the results 
of the preceding sections suggest that it may then be sufficient to compute 
only a few orders of scattering \citep[see e.g.][]{1971A&A....13....7H,
1980vandeHulst,2006Mishchenko}
instead of a more laborious complete multiple 
scattering calculation. This holds in particular when the optical thickness 
and/or the albedo of single scattering is not large. For example, for 
isotropic scattering in a homogeneous semi--infinite atmosphere with 
$a= 0.4$ and $\mu= \mu_0= 0.1$ the sum of the first two orders of 
scattering of the reflection function is only 1.6~$\%$ too low and 
the sum of the first three orders even less than 0.34~$\%$.

\section{Concluding remarks}
\label{sect8}

Numerous complicated equations occur in the theory of multiple light 
scattering in homogeneous and inhomogeneous plane--parallel atmospheres 
\citep[see e.g.][]{1950ratr.book.....C,1975QB603.A85S6213,
1980vandeHulst,1997Yan,2004Hovenier,2006Mishchenko}.
Since there is always a possibility that printed equations contain
errors and their derivations 
are not always given, it is useful to have simple checks available like 
letting $\mu$ and $\mu_0$ approach zero and comparing the results with 
expressions in this paper. 

On performing model computations it is usually very helpful to know and 
understand what happens with the reflection function or a related function 
in limiting cases
\citep[][]{1983Icar...55..187I}. This kind of knowledge is provided in 
this paper even for a complicated model of an inhomogeneous atmosphere 
with an arbitrarily reflecting surface underneath. In particular one should 
be prudent in the proximity of discontinuities like those presented in 
this paper.

\begin{figure}
\centering
\resizebox{8cm}{!}{\includegraphics{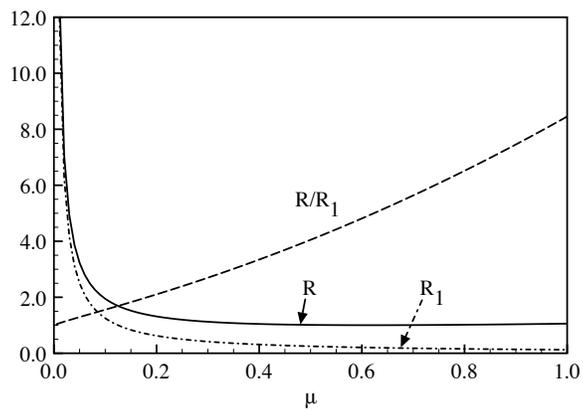}}
\caption{The reflection function ($R$) of a non--absorbing, homogeneous, 
         semi--infinite atmosphere with isotropic scattering in case 
         $\mu= \mu_0$ is plotted as a function of $\mu$. 
         Also shown are the contribution due to first order scattering 
         ($R_1$) and the ratio $R(\mu,\mu)/R_1(\mu,\mu)$, which in this 
         case equals $8 \mu R(\mu,\mu)$.}
\label{fig2}
\end{figure}

We have shown that a discontinuity exists for the reflection 
function which may hamper interpolation and extrapolation. This problem may 
be by-passed by multiplying the reflection function by a simple function of 
$\mu$ and $\mu_0$, like $\mu+\mu_0$ (cf.\ Eq.~\ref{eq9}), before performing 
the interpolation or extrapolation
\citep[][]{2000JQSRT..64..173K}. This is illustrated in 
Fig.~\ref{fig2} by the curve marked $R/R_1$ since in the case considered 
this curve equals $8 \mu R(\mu,\mu)$.

We have also shown that great care should be exercised with using 
approximative formulae for the reflection function, since they may lead 
to large errors for nearly horizontal directions. This holds, for instance, 
for the "rapid guess formula" and the very popular Lambert reflection law, 
which is often used, e.g.\ for cloudy atmospheres of planets 
\citep[see e.g.][]{2006Kok}.
Although approximative formulae exist that do not give 
large errors for nearly horizontal directions, more accurate results are 
possibly obtained by computing a few orders of scattering for such directions.


\bibliographystyle{aa}
\bibliography{0790AA_refs}

\clearpage
\newpage

\clearpage
\newpage

\end{document}